\documentclass[pra,twocolumn,amsmath,amssymb,nofootinbib,superscriptaddress,floatfix]{revtex4-2}

\pdfoutput=1
\usepackage[utf8]{inputenc}
\usepackage{mathtools}
\usepackage{bbold}
\usepackage{amsmath}
\usepackage{amssymb}
\usepackage{float}
\usepackage{braket}
\usepackage{xparse}
\usepackage{physics}
\usepackage{graphicx}
\usepackage{verbatim}
\usepackage{hhline}
\usepackage{bm}
\usepackage{color}
\usepackage[linesnumbered,ruled,vlined]{algorithm2e}
\SetKwInput{KwInput}{Input}                
\SetKwInput{KwOutput}{Output} 
\usepackage[toc,page]{appendix}
\usepackage[dvipsnames]{xcolor}
\usepackage{comment}

\SetCommentSty{mycommfont}

\newcommand{\beginsupplement}{%
        \setcounter{table}{0}
        \renewcommand{\thetable}{S\arabic{table}}%
        \setcounter{figure}{0}
        \renewcommand{\thefigure}{S\arabic{figure}}%
        \setcounter{equation}{0}
        \renewcommand{\theequation}{S\arabic{equation}}
     }

\begin{document}

\title{Exploring Finite Temperature Properties of Materials with Quantum Computers}
\author{Connor Powers$^*$}
\affiliation{Lawrence Berkeley National Lab, Berkeley, CA, USA}
\affiliation{University of Maryland, College Park, MD, USA\\$^*$cdpowers@umd.edu}
\author{Lindsay Bassman Oftelie}
\affiliation{Lawrence Berkeley National Lab, Berkeley, CA, USA}
\author{Daan Camps}
\affiliation{Lawrence Berkeley National Lab, Berkeley, CA, USA}
\author{Wibe A. de Jong}
\affiliation{Lawrence Berkeley National Lab, Berkeley, CA, USA}
\bibliographystyle{apsrev}
\raggedbottom
\begin{abstract}
Thermal properties of nanomaterials are crucial to not only improving our fundamental understanding of condensed matter systems, but also to developing novel materials for applications spanning research and industry. Since quantum effects arise at the nano-scale, these systems are difficult to simulate on classical computers. Quantum computers can efficiently simulate quantum many-body systems, yet current quantum algorithms for calculating thermal properties of these systems incur significant computational costs in that they either prepare the full thermal state on the quantum computer,  or they must sample a number of pure states from a distribution that grows with system size. Canonical thermal pure quantum (TPQ) states provide a promising path to estimating thermal properties of quantum materials as they neither require preparation of the full thermal state nor require a growing number of samples with system size. Here, we present an algorithm for preparing canonical TPQ states on quantum computers. We compare three different circuit implementations for the algorithm and demonstrate their capabilities in estimating thermal properties of quantum materials.  Due to its increasing accuracy with system size and flexibility in implementation, we anticipate that this method will enable finite temperature explorations of relevant quantum materials on near-term quantum computers.
\end{abstract}

\maketitle

\section{Introduction}
As the search for high-performance materials persists in research and industry alike, there is a need for better understanding the thermal properties of materials. In particular, exploring thermal properties of nanomaterials is of great interest, with applications ranging from energy production to nanoelectronics \cite{zhu2020generation,Zhang}. Quantum effects can dominate at the nano-scale, and the exponential growth of resources required to simulate quantum systems with classical computers makes the simulation of such materials quickly exceed the capabilities of even the largest classical supercomputers \cite{de2019massively}. Quantum computers, by contrast, are able to simulate quantum many-body systems efficiently \cite{lloyd1996universal,abrams1997simulations}. Therefore, quantum computers offer a promising route to studying thermal properties of quantum materials. While a plethora of simulations of systems at zero-temperature have been demonstrated on quantum devices in recent years \cite{Smith_2019, chiesa2019quantum, francis2020quantum, arute2020observation, Bassman_2020, yeter2021scattering, bassman2022constant, smith2022crossing}, the landscape of quantum algorithms to calculate finite temperature properties remains more sparse \cite{bassman2021computing, sun2021quantum, bassman2021simulating}. The main challenge in exploring finite-temperature properties on quantum computers lies in the preparation of thermal states.

Current quantum algorithms for thermal state preparation fall into two main categories. The first comprises algorithms that initialize the qubits into the full thermal (i.e., mixed) state. In this case, the thermal average of an observable can be computed directly by measuring the observable in this state. Examples include algorithms that prepare the Gibbs state using phase estimation \cite{Poulin,Riera_2012,DimensionReduction}, which require quantum circuits that are too large for near-term quantum devices, otherwise known as noisy intermediate-scale quantum (NISQ) computers \cite{preskill2018quantum}.  Other examples include the variational quantum thermalizer \cite{verdon2019quantum} and methods that prepare thermofield double states \cite{wu2019variational, zhu2020generation}, both of which rely on variational techniques. The variational nature of these algorithms necessitates the use of a cost function,  which generally becomes hard to compute as system size increases. Such methods are therefore difficult to scale to large or complex systems. Still other methods for generating the full thermal state require a number of ancilla qubits that scales with system size or complexity \cite{Terhal,Chowdhury2017}, thus limiting the size of systems that can be simulated on current quantum hardware.

Algorithms in the second category prepare an ensemble of pure states, one pure state at a time, where each pure state has been sampled according to the correct thermal distribution. Existing examples rely on Monte Carlo sampling techniques, Markov chains, or both \cite{Motta_QITE, QITE_dynamical, MeasurementBased, Temme_2011, Q2MA, Finite_Energies}. To calculate thermal averages, the desired observable is measured in each of the different pure states and the results are averaged over the ensemble.  As pure states are much easier to prepare on a quantum computer than mixed states, this model for thermal state preparation is more promising for NISQ computers. However, the number of samples required generally grows with the size of the system being simulated, which can lead to significant resource requirements for large systems.

Canonical thermal pure quantum (TPQ) states  \cite{Sugiura_2013} offer a promising way to estimate thermal averages on quantum computers. Thermal state approximation by TPQ states lies in a separate third category, as it neither incurs the quantum resources required to prepare a mixed state, nor relies on a number of samples that grows with system size.  TPQ states are formed by applying a specific non-unitary transformation, which is a function of the system Hamiltonian and inverse temperature, to a random state. The resulting state is shown to be representative of the thermal equilibrium in that observables measured in this state will approximate thermal averages. Remarkably, the error in the expectation value of the observable is bounded by an exponentially decreasing function of system size $N$. Thus, at sufficiently large $N$, the expectation value of an observable in only a single TPQ state will yield a very close approximation to the thermal average. The effectiveness of canonical TPQ states has been demonstrated classically \cite{Sugiura_2013}, but to our knowledge, has not been implemented on a quantum computer. 

Here, we present an algorithm for generating canonical TPQ states on quantum computers, enabling the estimation of finite temperature properties of materials on NISQ devices.  Our algorithm relies on a straightforward and scalable protocol for preparing the random state \cite{Richter_2021}, which allows for circuit depths to be tuned to find a balance between desired accuracy and feasibility of execution on NISQ hardware. Furthermore, the algorithm is agnostic to implementation of the non-unitary transformation of the random state, which can be tailored to the resource constraints of different quantum devices.  We compare three possible implementations for approximating the non-unitary transformation: (i) the quantum imaginary time evolution (QITE) algorithm \cite{Motta_QITE}, which can be better suited to devices constrained in qubit count, (ii) the dilated operator approach \cite{Probabilistic_Nonunitary}, which provides some advantages in resource requirements over the QITE algorithm when a single ancillary qubit is available, and (iii) the FABLE method \cite{FABLE} which is better suited for devices with limited coherence times. It is noted that Quantum Signal Processing may also be used to approximate this non-unitary operation, as discussed in Ref. \cite{coopmans2022predicting}.

We demonstrate our algorithm using each implementation for the Heisenberg model, a quintessential model used for studying a range of behaviors in materials \cite{billoni2005spin, gong2014emergent, jepsen2020spin, tanaka2020prediction, rodriguez2021turbulent}.  We anticipate that this algorithm will facilitate estimations of finite temperature properties of materials on near-term quantum devices.  Furthermore, since error in estimating thermal averages with TPQ states decreases with increasing system size, we believe this algorithm will only become increasingly useful as quantum hardware continues to grow in size in the coming years.

\section{Theoretical Framework} The (unnormalized) canonical TPQ state of a system of size $N$, governed by Hamiltonian $H$, at inverse temperature $\beta$ is defined as \cite{Sugiura_2013}
\begin{equation}
    \ket{\beta,N}=\hat{Q}\ket{\Psi_R}\equiv e^{-\beta H/2}\ket{\Psi_R}
\end{equation}
where $\ket{\Psi_R}=\sum_{i=1}^{2^N} c_i \ket{i}$ is a random state, defined by complex amplitudes ${c_i}$ which are uniformly sampled from the unit hypersphere such that $\sum_{i=1}^{2^N} |c_i|^2 = 1$, and $\ket{i}$ is an arbitrary orthonormal basis.  Defined this way, $\ket{\Psi_R}$ is characterized as a Haar-random pure state \cite{haarunitaries}.
We define 
\begin{equation}
    \langle\hat{A}\rangle_{\beta,N}^{ens} \equiv \frac{\Trace{[e^{-\beta H}\hat{A}]}}{\Trace{[e^{-\beta H}]}}
\end{equation}
as the ensemble expectation value of an operator $\hat{A}$ for a system of size $N$ at inverse temperature $\beta$, and
\begin{equation}
    \langle\hat{A}\rangle_{\beta,N}^{TPQ} \equiv \frac{ \bra{\beta,N}\hat{A}\ket{\beta,N}}{\bra{\beta,N}\ket{\beta,N}}
\end{equation}
as the corresponding expectation value of $\hat{A}$ in a single TPQ state. It is noted that $\hat{A}$ must be a low-degree polynomial of local operators for the following analysis to hold, but this group contains many prominently used observables including energy, magnetization, and a number of relevant correlation functions. The error between $\langle\hat{A}\rangle_{\beta,N}^{TPQ}$ and $\langle\hat{A}\rangle_{\beta,N}^{ens}$ is bounded by a value that becomes exponentially small with increasing system size $N$ \cite{Sugiura_2013}.  In practice, this means that for sufficiently large $N$, measuring the desired observable in a single TPQ state will provide a good approximation of the true thermal average.  Indeed, Ref. \cite{Sugiura_2013} found less than a $1\%$ error when estimating thermal properties using a single TPQ state for a quantum spin model with $N=30$.  At lower $N$, fidelity may be increased by averaging over the measured values from multiple TPQ states \cite{Sugiura_2013}. 

\begin{figure}[ht]
    \centering
    \includegraphics[width=0.99\linewidth]{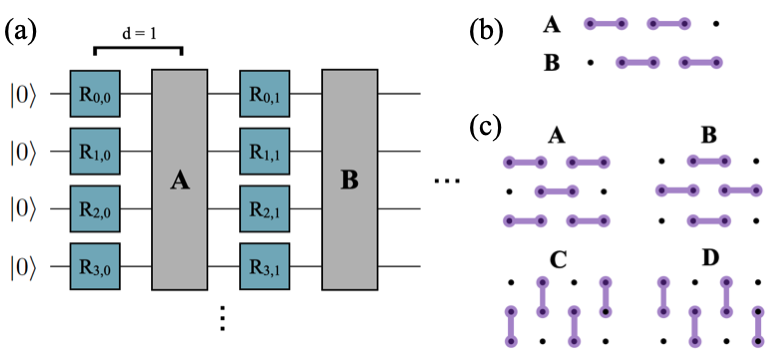}
    \caption{(a) Gate pattern for approximate Haar-random circuits. $R_{i,j}$ are randomly selected single-qubit rotation gates selected from a finite set with constraints (see main text). In 1D systems 2-qubit layers follow the pattern \textbf{ABAB...} with individual layer patterns specified in (b), while in 2D systems, 2-qubit layers follow the pattern \textbf{ABCDABCD...}, with individual layer patterns specified in (c).} 
    \label{fig:circuitpattern}
\end{figure}

The first step in preparing a TPQ state on a quantum computer is the preparation of $\ket{\Psi_R}$. We can approximate this Haar-random state using random quantum circuits constructed in the manner proposed in Ref. \cite{Richter_2021} and illustrated in Fig. \ref{fig:circuitpattern}.

The random circuits we consider are composed of ``blocks," where each block is composed of a layer of single-qubit rotation gates on every qubit followed by a layer of two-qubit entangling gates. The single-qubit rotation gates are selected from a finite set $\mathbb{A}=\{RX(\frac{\pi}{2}),RY(\frac{\pi}{2}),T\}$, with the constraint that no gate may be chosen for the same qubit two blocks in a row. In other words, if $R_{i,j}$ is the $j$th single-qubit gate acting on qubit $i$, then $R_{i,j} \in \mathbb{A}\setminus R_{i,j-1}$.

The two-qubit gate layers, as shown in Fig. \ref{fig:circuitpattern}b-c, follow a fixed pattern determined by the dimension of the Hamiltonian of interest; given a system dimension of $k$, there will be $2k$ different two-qubit gate layer patterns to loop through to ensure all coupling directions have been accounted for. In this way, the random circuits are easily generalizable to higher dimensions. For example, Fig.~\ref{fig:circuitpattern}b shows the two 2-qubit gate patterns that must be looped through for a one-dimensional (1D) system. Similarly, Fig.~\ref{fig:circuitpattern}c shows the four 2-qubit gate patterns to loop through for 2D systems. 

Random circuits of this form can be defined by a single parameter $d$, which sets the number of blocks and ultimately controls the circuit depth. Such circuits can be seen as approximating successively higher $t$-designs as $d$ is increased \cite{brandao2016local, tdesign_ref,Nakata_2017, Boixo_2018}.  (A unitary $t$-design is an approximation of a Haar random unitary which accurately simulates the first $t$ moments of Haar
random unitaries.  Higher order $t$-designs generate states whose properties converge to those of Haar-random states). We can see this convergence by plotting the entropy of the resulting random state.  This plot is useful for determining how large $d$ needs to be for a given system size; the necessary $d$ will be the point at which the state entropy is sufficiently converged to the value $\ln{N}-1+\gamma$ characteristic of Haar-random states, where $\gamma \approx 0.577$ is the Euler-Mascheroni constant. See Section I in the Supplementary Information (SI) for these convergence plots with and without simulated device noise. Here, we set $d=20$, unless otherwise specified.

Once the random state $\ket{\Psi_R}$ has been generated, the non-unitary operator $\hat{Q}$ must be applied to generate a canonical TPQ state. Given that quantum computers can only perform unitary operations, this must be implemented by a unitary approximation of the non-unitary transformation.  Several approximation methods exist and in practice, the method choice is informed by available quantum resources. Here, we demonstrate our algorithm with three different approaches, including QITE \cite{Motta_QITE}, a dilated operator approach \cite{Probabilistic_Nonunitary}, and the FABLE method \cite{FABLE}. 

With QITE, neither ancillary qubits nor post-selection of results are required.  This however, comes at the cost of deeper circuits that require significant classical resources to generate. Circuit generation time with QITE grows quickly as the simulated system size is increased, though this can be mitigated by decreasing the domain-size chosen to generate the QITE approximation.  When average correlation lengths of the simulated system are small, a domain-size significantly smaller than total system size can be chosen to dramatically decrease the computational complexity of generating the QITE circuit.  However, this can result in reduced accuracy when correlation lengths of the system are larger than the selected domain size.  

The unitary dilation method requires a single ancilla qubit as well as post-selection of results.  Any experiment in which the ancilla qubit is measured to be in the `1' state must be discarded, which increases the number of shots required to generate good results.  Overall, the unitary dilation method tends to generate shallower circuits than QITE, but at the price of requiring more shots. 

Finally, FABLE requires a register of ancilla qubits that grows linearly with the simulated system size as well as post-selection of results.  For a system size of $N$, FABLE requires $N+1$ ancilla qubits, all of which must be measured to be in the `0' state (otherwise the result must be discarded).  While its post-selection requirement based on such a large ancilla register will necessitate a very large number of shots, FABLE offers the most favorable circuit depths and circuit generation times of the three methods.

\begin{figure}[h]
    \centering
    \includegraphics[width=0.99\linewidth]{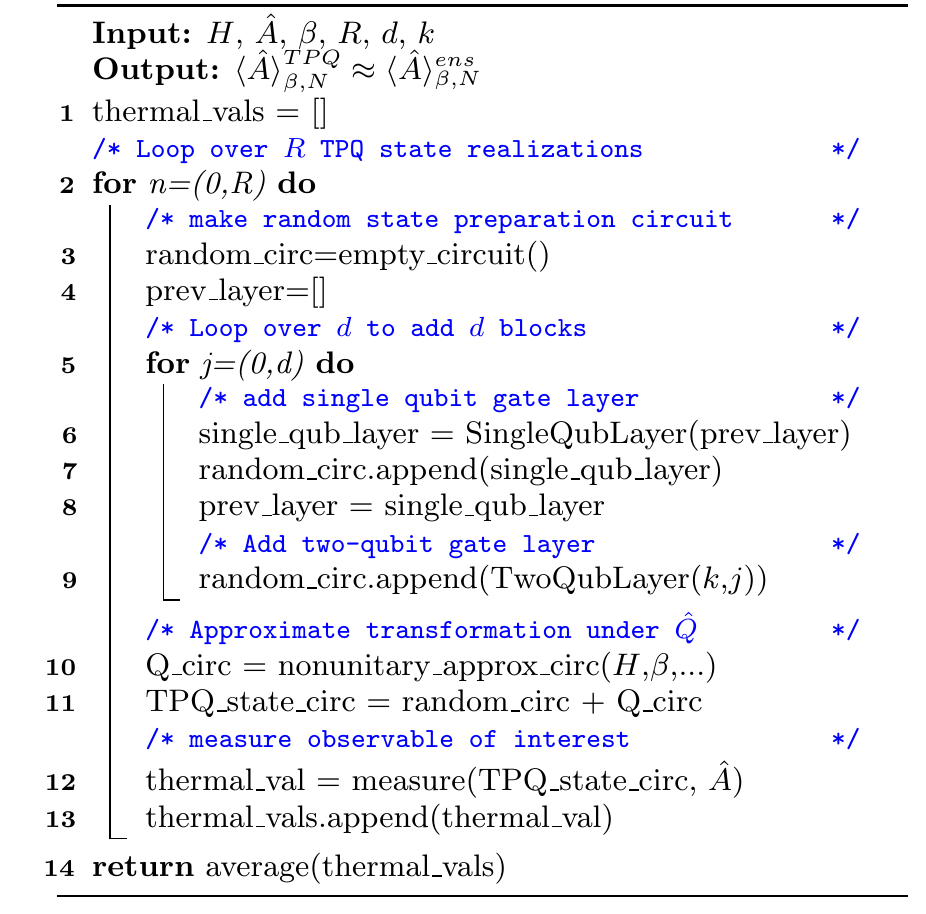}
    \caption{Pseudocode algorithm for computation of thermal averages with canonical TPQ states on quantum computers.}
    \label{fig:generalalg}
\end{figure}

The algorithm shown in Fig. \ref{fig:generalalg} summarizes our method for approximating thermal averages with TPQ states on quantum computers.  First, the $d$-layer circuit is constructed to generate $\ket{\Psi_R}$.  Layers of randomly chosen single-qubit rotation gates are alternated with layers of 2-qubit entangling gates following a pattern set by the Hamiltonian dimension $k$. Next, a unitary approximation of the non-unitary transformation $\hat{Q}$ is applied. The algorithm is flexible in how this non-unitary operation is implemented. The observable of interest is then measured in the resulting canonical TPQ state, and this process is repeated $R$ times to construct and measure $R$ distinct TPQ states. Finally, if applicable, the average of the measured thermal values is taken.

If QITE is chosen to implement line 10 in Fig. \ref{fig:generalalg}, the non-unitary operator $\hat{Q}$ is viewed as an evolution through imaginary time.  To find a unitary approximation, this evolution is broken up into time-steps of imaginary time, and each time-step is approximated by a unitary sub-circuit.   Increasing the number of time-steps into which the imaginary-time evolution is broken will increase the transformation fidelity, i.e., the accuracy of the approximation.  However, the size of the quantum circuit grows linearly with the number of time-steps.  The sub-circuit for each time-step is constructed based on a so-called domain-size, which can be set to be equal to or less than the simulated system size, and should be chosen based on the average correlation length within the simulated system. The larger the domain-size the more accurate the QITE approximation will be, though this comes at the cost of increased computation complexity in generating the sub-circuits.  When using QITE, the number of imaginary time-steps and the domain-size must be chosen appropriately by the user for the simulated system.

If line 10 in Fig. \ref{fig:generalalg} is instead implemented with the dilated operator approach, an ancillary qubit initialized in state $\ket{1}$ must be added to create the augmented input state $\rho_{in} = (\ket{1}\bra{1}) \otimes \rho_R$, where $\rho_R = \ket{\Psi_R}\bra{\Psi_R}$. A dilated unitary operator $\hat{\Omega}$ is then constructed in the new $2^{N+1}$ dimensional Hilbert space that will approximate the action of $\hat{Q}$ on the original $2^N$ dimensional system Hilbert space \cite{Probabilistic_Nonunitary}.  $\hat{\Omega}$ is defined in terms of the non-unitary operator $\hat{Q}$ as
\begin{equation}
    \hat{\Omega} \equiv \exp(i\epsilon \begin{pmatrix} 0&-i\hat{Q}\\i\hat{Q}^\dagger&0 \end{pmatrix})
\end{equation}
where $\epsilon$ is a parameter that controls the performance of the operator. The dilated unitary operator $\hat{\Omega}$ is then applied to the augmented initial state $\rho_{in}$, and the ancillary qubit is measured. If it is measured in the state $\ket{0}$, the original $N$-qubit system of interest has been successfully transformed by an approximation to the non-unitary operator $\hat{Q}$. The probability that the ancillary qubit is measured in state $\ket{0}$, also called the probability of success, is denoted $P_0$. If the ancillary qubit is measured in state $\ket{1}$, the results of the entire circuit are discarded. \cite{Probabilistic_Nonunitary}.

To help quantify the performance of this dilated operator, it is useful to define a fidelity metric that measures how close the effective transformation on the $N$-qubit system is to the desired non-unitary transformation under $\hat{Q}$. This fidelity can be defined as:
\begin{equation}
    F = \Trace{\sqrt{\sqrt{\rho_0} \ket{\beta,N}\bra{\beta,N}\sqrt{\rho_0}}}.
\end{equation}
Here, $\rho_0$ is the density matrix of the $N$-qubit system after tracing out the ancillary qubit, provided that it is measured in the state $\ket{0}$.
The probability of success $P_0$ and the transformation fidelity $F$ are ultimately controlled by the user's choice of $\epsilon$.  While a small $\epsilon$ is required to obtain an accurate approximation to the non-unitary operator (i.e. high transformation fidelity $F$), decreasing $\epsilon$ generally decreases the probability of success $P_0$, which means more shots need to be run until the ancilla is measured in the $\ket{0}$ state, indicating a successful transformation. See Section II of the SI for more information regarding the behavior of $F$ and $P_0$ with varying $\epsilon$.  

While this method introduces a single ancillary qubit, its advantage is that increasing the accuracy of the approximate non-unitary transformation only requires increasing the average number of shots needed to see achieve a successful transformation. This stands in contrast to the QITE algorithm, where increasing the accuracy of the approximation requires a larger number of imaginary time-steps and subsequently deeper circuits. If the dilated operator approach is preferred, then the only change in Fig. \ref{fig:generalalg} is the addition of an ancillary qubit initialized in the $\ket{1}$ state which is not included in the random circuit construction. The non-unitary transformation approximation step then takes in the additional argument $\epsilon$.

The third approach we follow to implement line 10 in Fig. ~\ref{fig:generalalg} is a \emph{block-encoding} of the non-unitary operator $\hat Q$, which
is the embedding of the non-unitary in the leading principal block of a larger
unitary $U$,
\begin{equation}
U = \begin{pmatrix} \hat Q & *\, \\ * & *\, \end{pmatrix}.
\label{eq:BE}
\end{equation}
Here, $*$ indicate arbitrary matrix elements.  Assuming $a$ ancilla qubits are used to block encode the $2^N$ dimensional $\hat Q$, the operator $U$ is constructed in the $2^{N+a}$ dimensional Hilbert space.  Next, we apply $U$ to the augmented input state $\rho_{\text{in}} = (\ket{0}^{\otimes a} \bra{0}^{\otimes a}) \otimes \rho_R$. If all $a$ ancillary qubits are measured in the $\ket{0}$ state, the $N$-qubit operator $\hat Q$ has been successfully applied
to $\rho_R$. The likelihood of a successful measurement can be increased through
amplitude amplification.

We use FABLE~\cite{FABLE} to generate a circuit for Eq. \ref{eq:BE} based on the non-unitary operator $\hat Q$. FABLE circuits require $N+1$ ancilla qubits to encode an $N$-qubit operator and are fast to generate for small to medium-sized problems as the circuit generation algorithm scales as $\mathcal{O}(N 4^N)$.  The advantage of FABLE over the dilated operator approach is that no performance parameter $\epsilon$ is required and the circuits can be generated more efficiently while requiring fewer CNOT gates. The main disadvantage is that significantly more ancillary qubits are required.

In Tables \ref{table1}-\ref{table3}, we compare the three different methods utilized in this work for implementing the non-unitary evolution step of TPQ state preparation on quantum computers. Specifically, the three methods are compared across three metrics relevant to practical implementations on near-term devices: CNOT gate count of involved circuits (Table \ref{table1}), ancillary qubit requirements (Table \ref{table2}), and the time to classically generate the required circuits (Table \ref{table3}).  All data is generated for simulation of $N$-spin Heisenberg models (described in more detail in the following section).  We first attempted to generate the QITE circuits with the XACC software package \cite{xacc_2020}, however, this code requires the domain-size to be set equal to the system size, which made going to a system size of $N=4$ prohibitively expensive for our chosen model. For $N > 3$, we therefore used the ArQTiC software package \cite{bassman2021arqtic} to generate the QITE circuits.  However, ArQTiC only allows a maximum domain-size of 3.  Therefore, for systems with $N > 3$, a truncated domain-size of 3 had to be used, resulting in reduced accuracy of results.  In Tables \ref{table1}-\ref{table3}, results from XACC are provided in the column labeled `QITE', while results from ArQTiC are provided in the column labeled `Inexact QITE', since we expect these results to have lowered accuracy due to domain-size truncation \cite{Motta_QITE}. Since gate counts and/or circuit generation times of the QITE and dilated operator methods may vary between circuit realizations even for a given system size, data shown for these methods have been averaged over 10 circuit realizations.  All data is collected for an inverse temperature of $\beta=1$, and for QITE data, the imaginary time evolution operator is broken into 10 imaginary time steps.

\begin{table}[h]
\begin{center}
\begin{tabular}{ |c||c|c|c|c| } 
\multicolumn{5}{c}{CNOT Count}\\ 
\hline
\noalign{\vskip 2mm}
 \multicolumn{1}{c}{N}&\multicolumn{1}{c}{QITE}&\multicolumn{1}{c}{Inexact QITE}&\multicolumn{1}{c}{Dilated Operator}&\multicolumn{1}{c}{FABLE}\\
 \hhline{-||----}
 2 & $14$ &$20$ & $41$ & $16$ \\ 
 3 & $97$ &$963$& $218$ & $64$ \\ 
 4 & - &$1957$ & $1025$ & $256$ \\ 
 5 & -&$2945$ & $4474$ & $1024$ \\ 

\hhline{-||----}
\end{tabular}
\end{center}
\caption{CNOT gate count at various system sizes $N$ for approximating the imaginary time evolution step of TPQ state preparation with the QITE, inexact QITE, dilated operator, and FABLE algorithms at $\beta=1$.}
\label{table1}
\vspace{0.5cm}
\begin{center}
\begin{tabular}{ |c||c|c|c|c| } 
\multicolumn{5}{c}{Ancillary Qubits}\\ 
\hline
\noalign{\vskip 2mm}
 \multicolumn{1}{c}{N}&\multicolumn{1}{c}{QITE}&\multicolumn{1}{c}{Inexact QITE}&\multicolumn{1}{c}{Dilated Operator}&\multicolumn{1}{c}{FABLE}\\
 \hhline{-||----}
 2 & 0 & 0 & 1 & 3 \\ 
 3 & 0 & 0 & 1 & 4 \\ 
 4 & 0 & 0 & 1 & 5 \\ 
 5 & 0 & 0 & 1 & 6 \\ 

\hhline{-||----}
\end{tabular}
\end{center}
\caption{Ancillary qubit requirement at various system sizes $N$ for approximating the imaginary time evolution step of TPQ state preparation with the QITE, inexact QITE, dilated operator, and FABLE algorithms at $\beta=1$.}
\label{table2}
\vspace{0.5cm}

\begin{center}
\begin{tabular}{ |c||c|c|c|c| } 
\multicolumn{5}{c}{Circuit Generation Time [$s$]}\\ 
\hline
\noalign{\vskip 2mm}
 \multicolumn{1}{c}{N}&\multicolumn{1}{c}{QITE}&\multicolumn{1}{c}{Inexact QITE}&\multicolumn{1}{c}{Dilated Operator}&\multicolumn{1}{c}{FABLE}\\
 \hhline{-||----}
 2 & $2.85$ &$0.719$ & $0.970$ & $2.14\text{x}10^{-3}$ \\ 
 3 & $1.44\text{x}10^2$ &$3.71$ & $1.53$ & $6.39\text{x}10^{-3}$ \\ 
 4 & -  &$7.53$ & $4.44$ & $2.61\text{x}10^{-2}$ \\ 
 5 & -  &$11.51$ & $17.4$ & $8.71\text{x}10^{-2}$ \\ 
\hhline{-||----}
\end{tabular}
\end{center}
\caption{Wall-clock circuit generation time at various system sizes $N$ for approximating the imaginary time evolution step of TPQ state preparation with the QITE, inexact QITE, dilated operator, and FABLE algorithms at $\beta=1$.}
\label{table3}
\end{table}

As seen in Tables \ref{table1}-\ref{table3}, while the QITE algorithm has a clear advantage in that it requires no ancillary qubits, its scaling of circuit generation time is seen to be practically prohibitive even at relatively small system size. Therefore, this method seems to be best suited for simulating very small systems on equally small near-term devices. Switching to an inexact QITE with the domain-size truncated to 3 significantly alleviates the circuit generation time scaling with system size at the cost of increased entangling gate counts and reduced accuracy of results. The dilated operator method appears to bridge the resource requirement gaps of QITE and FABLE methods; it requires a minimal ancillary qubit requirement that is independent of simulated system size, and has more favorable circuit generation time scaling than QITE, but at a cost of increased CNOT gate counts. For small systems, the latter may be mitigated by some circuit synthesis techniques such as QFAST \cite{QFAST} and LEAP \cite{LEAP}. If sufficient ancillary qubit resources are available, the FABLE technique displays the best CNOT count and circuit generation time scalings of the three methods. Therefore, it is the best option when coherence times are a dominant limitation on experiments and ample qubit counts are available.

\section{Demonstration}
We now demonstrate our method by calculating the thermal energy of a Heisenberg spin model under an external magnetic field at various system sizes and inverse temperatures. The Hamiltonian of an $N$-spin Heisenberg model is given by
\begin{equation}
    \hat{H} = \sum_\alpha \sum_{\langle i,j \rangle} J_{\alpha,ij} \sigma^\alpha_i\sigma^\alpha_{j} + h_x\sum_{i=1}^{N} \sigma^x_i
    \label{XYZ_Hamiltonian}
\end{equation}
where $J_{\alpha,ij}$ ($\alpha \in \{x,y,z\}$) gives the strength of the exchange coupling interaction between nearest neighbor spin pair $\langle i,j \rangle$ in the $\alpha$-direction, $h_x$ is the strength of an externally applied magnetic field in the $x$-direction, and $\sigma^\alpha$ are Pauli matrices. For results presented in this work, we set $J_x = 0.5$, $J_y = 1.25$, $J_z = 2.0$, and $h_x = 1.0$. 

From TPQ state formalism it is known that, on average, the squared error from estimating a thermal value with a single TPQ state is bounded by a function that becomes exponentially small with increasing simulated system size.~\cite{Sugiura_2013} At small system sizes, some of this error may be mitigated through averaging over multiple TPQ state realizations. For more details and numerical results, see Section III in the SI.

\begin{figure}[ht]
    \includegraphics[width=0.9\linewidth]{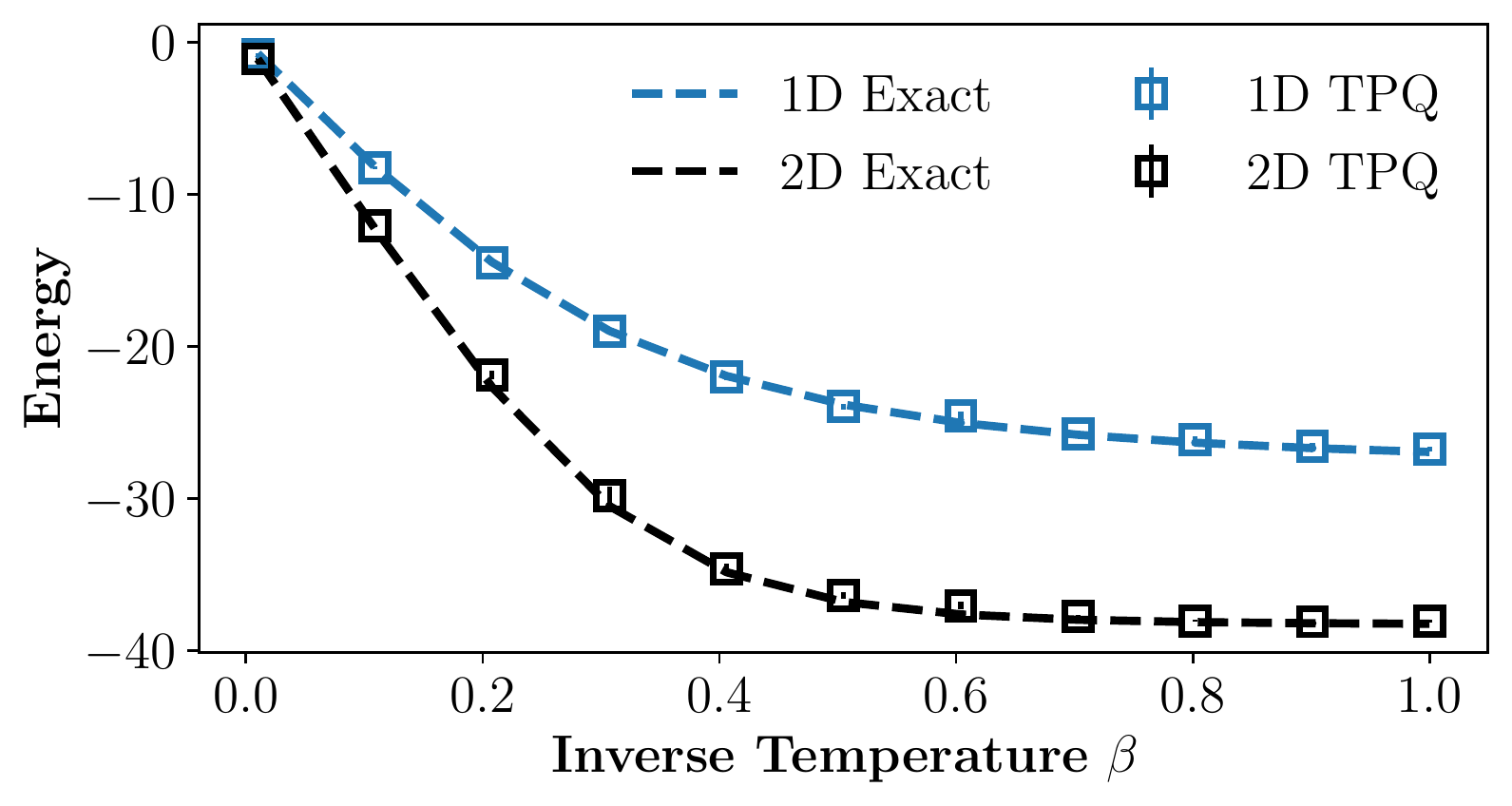}
    \caption{Thermal energy, as a function of inverse temperature $\beta$ of a 1D Heisenberg model on a 12-qubit lattice (blue) and a 2D Heisenberg model on a 4x3 lattice (black). Shown are results averaged over 10 TPQ state realizations. The true ensemble thermal energies (dashed) are plotted for reference.}
    \label{fig:betasweep}
\end{figure}

In Fig. \ref{fig:betasweep}, we present numerical results demonstrating the efficacy of the method in calculating thermal energies of 12-spin 1D and 2D Heisenberg models at various inverse temperatures. The random circuits are built through the aforementioned protocol and the non-unitary operation $\hat{Q}$ is numerically simulated.  The thermal energies in both the 1D and 2D systems are closely approximated by averaging over just 10 canonical TPQ states. 

\begin{figure}
    \includegraphics[width=0.95\linewidth]{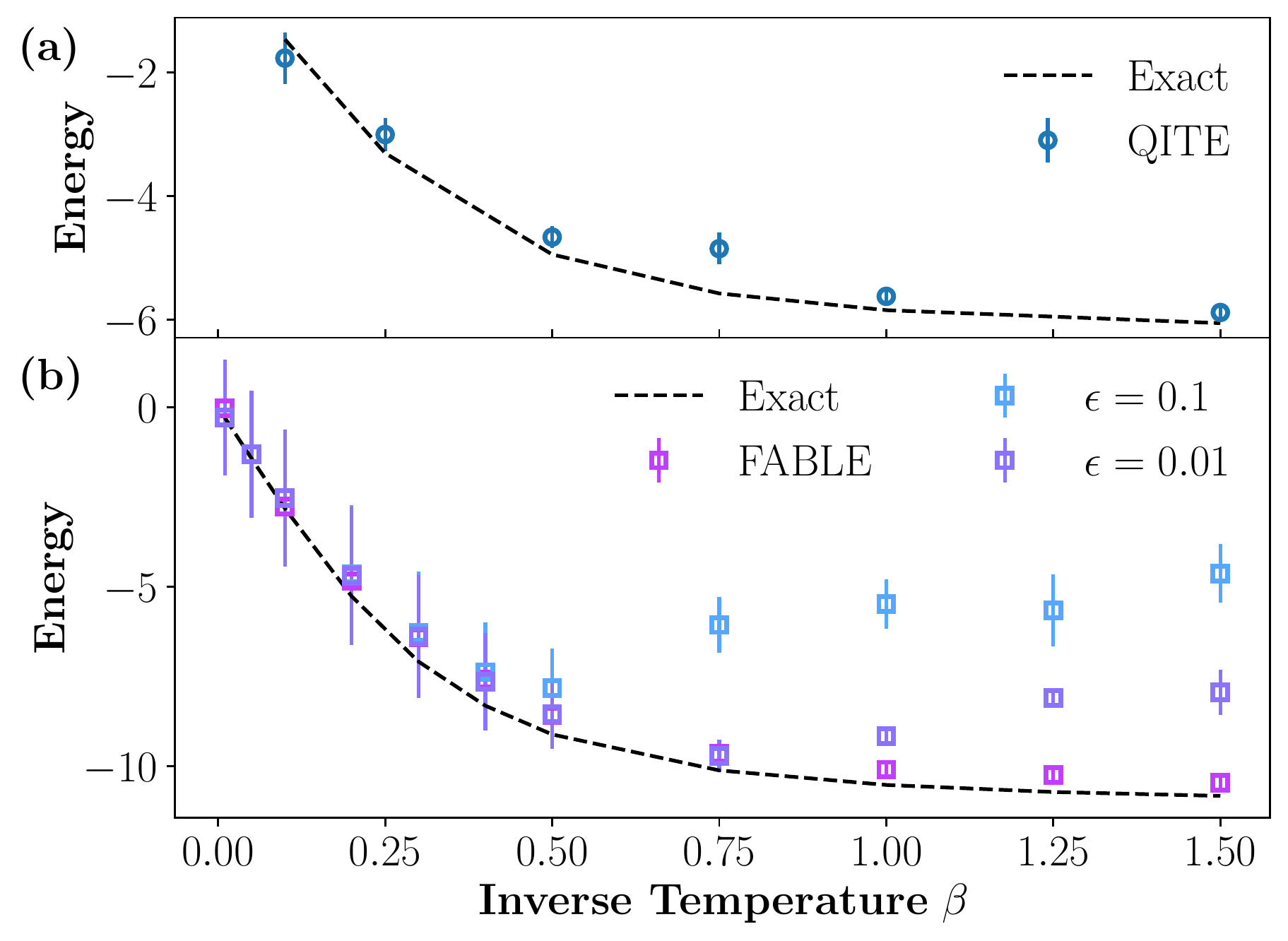}
    \caption{Thermal energy, as a function of inverse temperature $\beta$, calculated with TPQ states on a noiseless quantum simulator for a (a) 3-qubit 1D Heisenberg model using the QITE algorithm to approximate $\hat{Q}$ (b) 5-qubit 1D Heisenberg model using FABLE and a dilated operator approach to approximate $\hat{Q}$.}
    \label{fig:simresults}
\end{figure}

Next, we demonstrate quantum simulator results of our method using the three different quantum circuit implementations of $\hat{Q}$. In Fig. \ref{fig:simresults}a, the thermal energy of a 3-qubit 1D Heisenberg model as a function of inverse temperature $\beta$ is computed using QITE to approximate $\hat{Q}$. The QITE circuit was generated using XACC \cite{xacc_2020}. Results are averaged over $R=10$ TPQ states, with error bars showing the uncertainty given by one standard deviation divided by $\sqrt{R}$. In Fig. \ref{fig:simresults}b, the thermal energy of a 5-qubit 1D Heisenberg model as a function of inverse temperature is approximated using the dilated operator and FABLE approaches to implement $\hat{Q}$. Results using two different values of the dilated operator parameter $\epsilon$ are included to demonstrate its impact on performance.  While smaller $\epsilon$ leads to significantly better results for larger $\beta$, it requires many more executions as the success probability decreases with smaller $\epsilon$.  After constructing the requisite dilated operator to approximate the non-unitary transformation $\hat{Q}$, Qiskit was utilized to decompose it into a circuit with the basis gate set of the IBM's ``ibmq$\_$brooklyn" device. Presented results are averaged over $R=10$ distinct TPQ states, and the error bars show uncertainty.

The dilated operator method appears to be consistently accurate at low $\beta$, while its performance begins degrading past some threshold inverse temperature. This threshold can be increased by decreasing the performance parameter $\epsilon$. Results derived from using the FABLE technique have comparable accuracy to the dilated operator results at high temperatures and do not appear to exhibit a similar degradation at low temperatures. 

\begin{figure}
    \centering
    \includegraphics[width=0.95\linewidth]{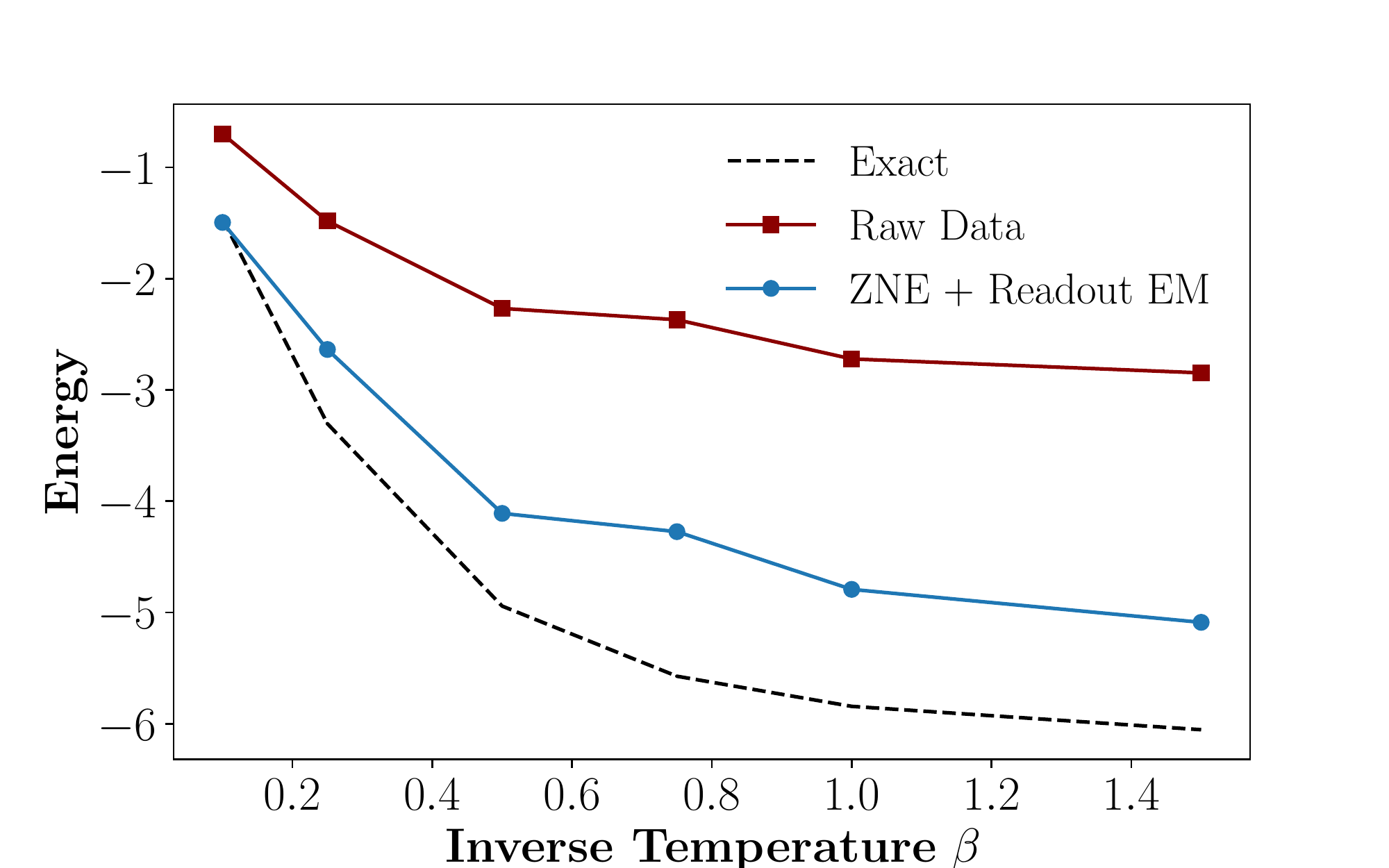}
    \caption{Thermal energy data from canonical TPQ states, as a function of inverse temperature $\beta$, of a 3-qubit 1D Heisenberg model executed on IBM's ``ibmq$\_$brooklyn" quantum computer. The QITE algorithm was performed with 10 imaginary time-steps for each value of $\beta$. Raw hardware data is presented in comparison to results after readout error mitigation (EM) and zero-noise extrapolation (ZNE) have been applied.}
    \label{fig:hardware}
\end{figure}

Finally, Fig. \ref{fig:hardware} demonstrates results from circuits run on IBM's ``ibmq$\_$brooklyn" quantum computer. Using the QITE version of the algorithm, we calculated the thermal energy of a 3-site Heisenberg model as a function of inverse temperature $\beta$ while employing readout error mitigation (EM) and zero-noise extrapolation~\cite{zne1,zne2} (ZNE) techniques to reduce error. The QITE circuits were compressed through numerical optimization using the  QSearch~\cite{qsearch} tool. 
Fig. \ref{fig:hardware} shows that while raw hardware data from generating and utilizing canonical TPQ states to measure thermal values has a significant amount of noise, this can be mitigated by standard techniques such as readout EM and ZNE. These error mitigation techniques can effectively accelerate the timeline in which this method for calculating thermal averages is achievable on near-term devices.
\section{Conclusion} We have presented a new method for approximating finite temperature properties of materials on quantum computers using TPQ states.  We demonstrated its efficacy through approximating thermal energies of Heisenberg models in one and two dimensions. To demonstrate flexibility in how the non-unitary step of the algorithm is implemented, we presented results from a quantum simulator derived from performing this transformation with the QITE algorithm, a dilated operator approach, and the recently-developed FABLE method. We also present hardware results executed on IBM's ``ibmq$\_$brooklyn" quantum computer, showing the efficacy of the method when combined with standard error mitigation techniques. Due to the increasing accuracy of thermal observables derived from TPQ states with system size, as well as the flexibility in the implementation of our algorithm, we expect our method for computing thermal properties of materials to be increasingly useful as higher quality qubits continue to be added to current quantum computers as we progress through the NISQ era and beyond.

\section*{Supplementary Material}
\beginsupplement
\subsection{Haar-random behavior of random circuit states}
A useful characteristic of a random state $\ket{\psi_R}$ is its entropy $H$:
\begin{equation}
    H_{\psi_R}=-\sum_{k=1}^{2^N} p_k \ln(p_k)
\end{equation}
where $p_k \equiv |\bra{k}\ket{\psi_R}|^2 = |c_k|^2$ \cite{Richter_2021}.

In a true Haar-random state, $p_k$ values follow the Porter-Thomas distribution, which can be shown to lead to $H_{Haar} = \ln(2^N)-1+\gamma$ where $\gamma \approx 0.577$ is Euler's constant \cite{Boixo_2018}.
\begin{figure}[ht]
    \centering
    \includegraphics[width=0.95\linewidth]{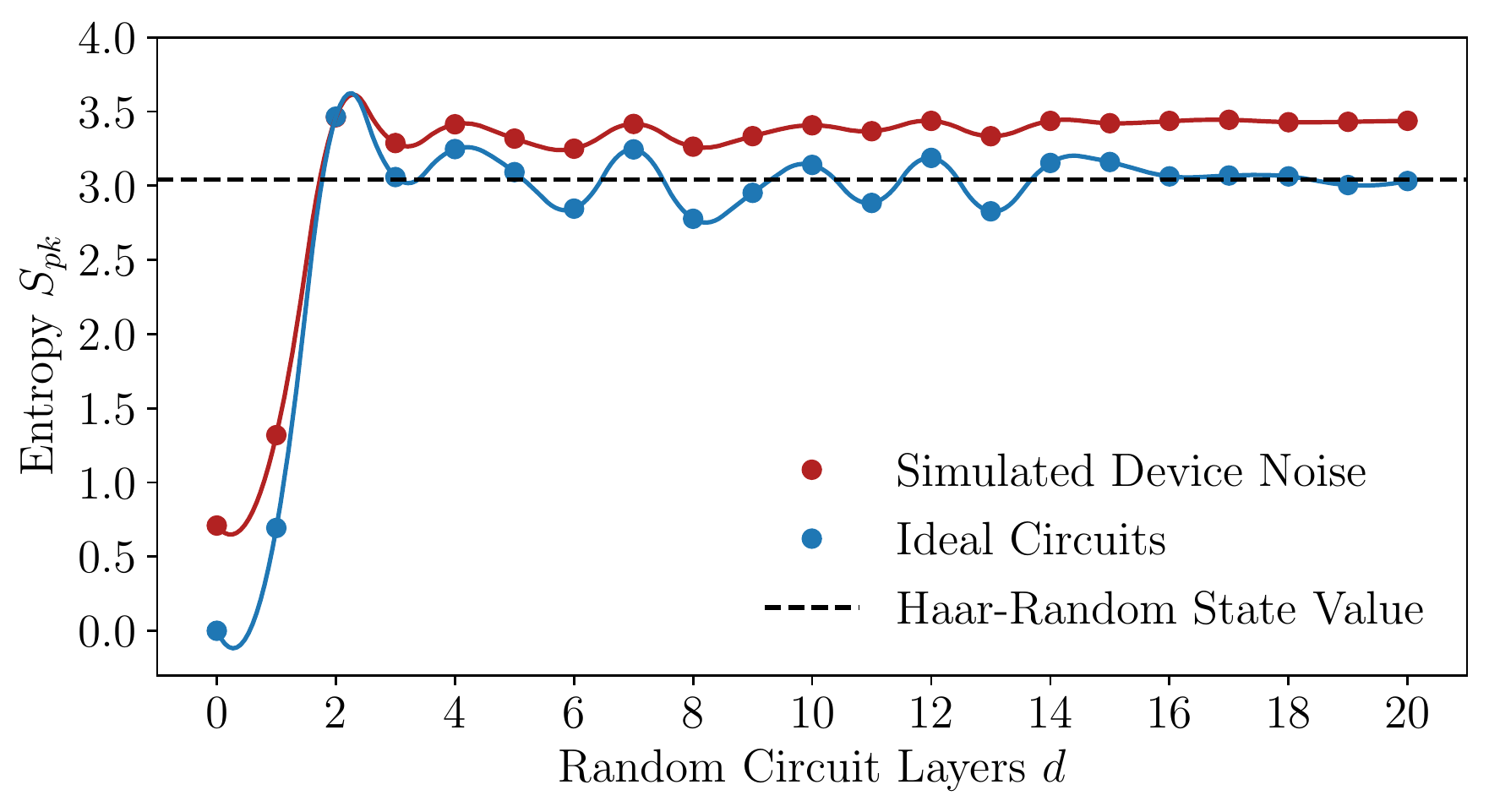}
    \caption{Entropy in 5-qubit random states prepared by random circuits of increasing depth $d$ with and without simulated device noise mimicking the IBM "Quito" quantum computer. A B-spline of degree 2 is fit to the data for readability. The entropy of ideal circuits approaches the value expected in a 5-qubit Haar-random state at $d \leq 10$, in line with the findings of Ref. \cite{Richter_2021}, while the simulated device noise causes the entropy to approach a slightly higher value.}
    \label{fig:entropy_buildup}
\end{figure}

As shown in Fig. \ref{fig:entropy_buildup}, the entropy of the states generated by this random circuit structure approaches what is characteristic of Haar-random states in relatively few layers when circuits are ideally simulated. The two-qubit gates can be either CZ or CNOT gates, but since CZ sets have shown to converge upon Haar-random values slightly faster than CNOT sets \cite{Richter_2021}, results presented here utilize CZ gates. We also illustrate the effects of simulated device noise; the entropy of the $N$-qubit random state generated with noisy circuits converges to a value slightly higher than what would be expected of an $N$-qubit Haar-random state. This is not anticipated to meaningfully affect resulting thermal value estimations.

\subsection{Dilated operator fidelity and probability behavior}
\begin{figure}[ht]
    \centering
    \includegraphics[width=0.99\linewidth]{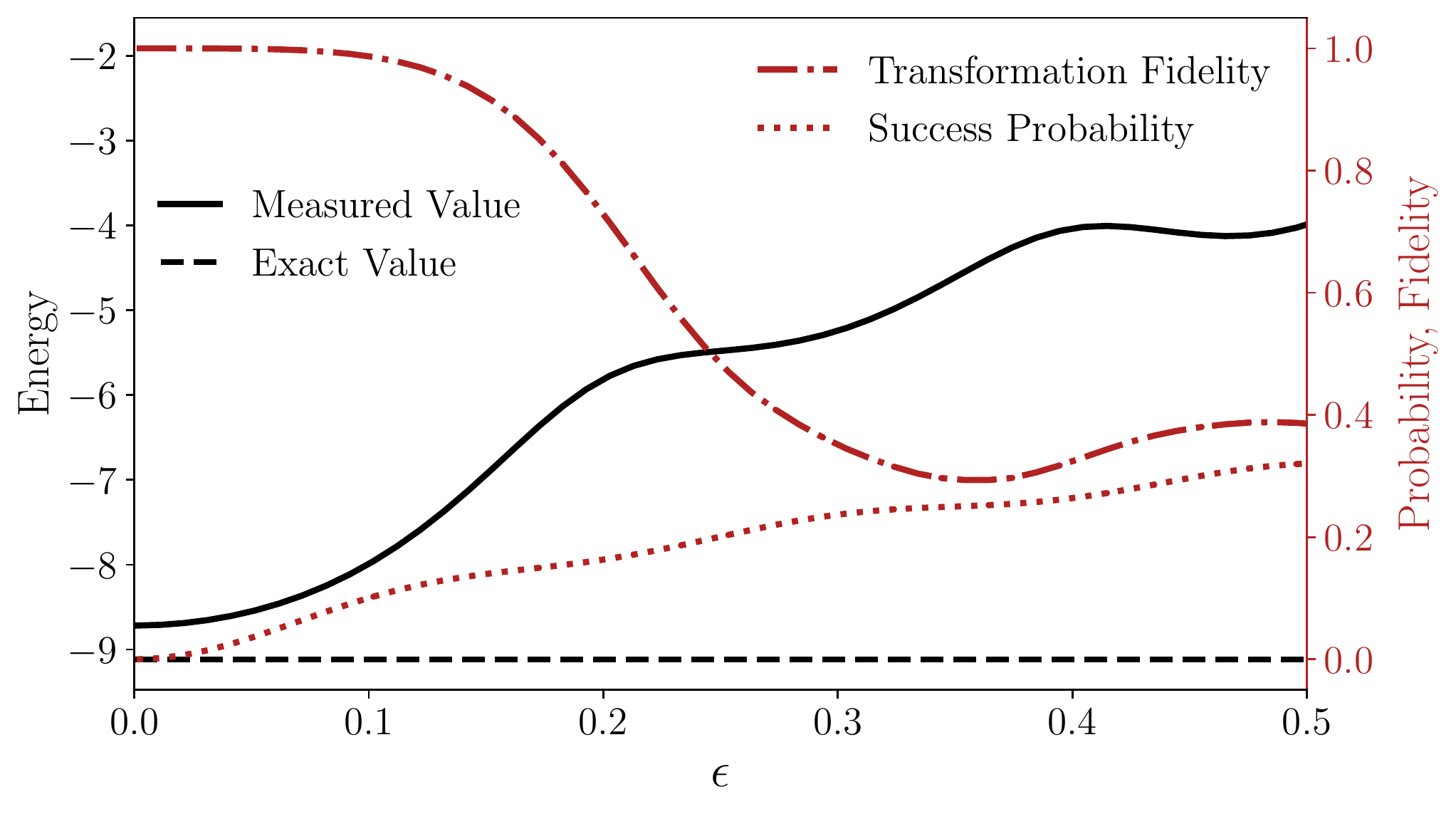}
    \caption{Average measured energy, probability of transformation success $P_0$, and transformation fidelity $F$ of $R=100$ TPQ states realized via the dilated operator approach for varying $\epsilon$ at $N=5$, $J_x = 0.5$, $J_y = 1.25$, $J_z = 2.0$, $h_x = 1.0$, and $\beta=0.5$. The ensemble value is shown for reference.}
    \label{fig:probabilistic}
\end{figure}

As mentioned in the main text, the performance of the dilated operator approach to approximating $\hat{Q}$ exhibits complex dependencies on the parameter $\epsilon$. The main performance metrics considered are the fidelity of transformation and the probability of a successful transformation occurring, defined as $F$ and $P_0$ in the main text respectively, so it is useful to examine how these properties change with $\epsilon$. As an example task, we will seek to determine the thermal energy of a 3-site Heisenberg model with $J_x = 0.5$, $J_y = 1.25$, $J_z = 2.0$, and $h_x = 1.0$ at inverse temperature $\beta=0.5$. Fig. \ref{fig:probabilistic} shows the thermal energy $\langle H \rangle^{TPQ}_{\beta,H}$ averaged over $R=100$ independent TPQ states realized using the dilated operator approach for varying $\epsilon$, along with the probability of success $P_0$, transformation fidelity $F$, and the true ensemble thermal energy $\langle H \rangle^{ens}_{\beta,H}$.

As seen in Fig. \ref{fig:probabilistic}, the transformation fidelity is highest at low $\epsilon$, while this is where the probability of success is lowest. Therefore, one must pick a value of $\epsilon$ such that the transformation fidelity is high enough to recover a good enough approximation of the observable of interest, while making sure the probability of success is large enough to be practical. This threshold will depend on how many shots are feasible to run in a particular research setting.

\subsection{TPQ State Error Trends}

\begin{figure}[H]
    \centering
    \includegraphics[width=0.4\textwidth]{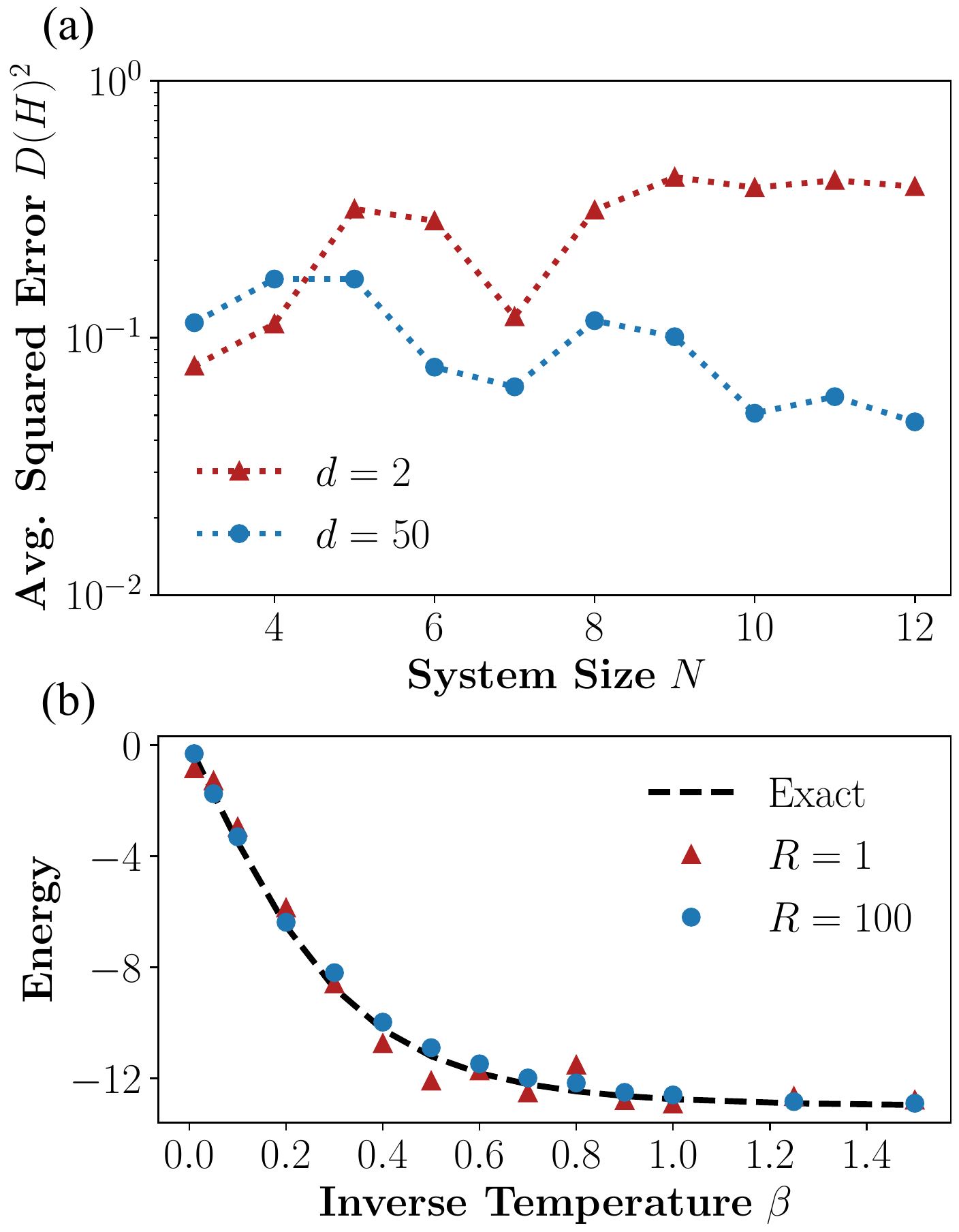}
    \caption{(a) Squared error of thermal energy calculated from TPQ states with $d=2$ and $d=50$ as a function of system size $N$ for a 1D Heisenberg model at inverse temperature $\beta=0.5$. Results are averaged over 100 TPQ state realizations. (b) Comparing thermal energies of a 6-spin Heisenberg model as a function of inverse temperature $\beta$ calculated from a single TPQ state and 100 TPQ states (R=1 and R=100 respectively). Exact values are shown for comparison.}
    \label{fig:errorfig}
\end{figure}

Fig. \ref{fig:errorfig}a demonstrates observed error trends with increasing system size $N$, averaged over results from $R=100$ distinct TPQ states, for a 1D Heisenberg model at inverse temperature $\beta = 0.5$. We define the average squared error as $D(H)^2=\overline{(\langle \hat{H} \rangle_{TPQ}-\langle \hat{H} \rangle_{ens})^2}$ in alignment with the TPQ state formalism laid out in Ref. \cite{Sugiura_2013}. Results are shown from TPQ states formed from random circuits with $d=2$, when the approximation to a Haar-random state is expected to be poor, and $d=50$, when this approximation is expected to be close. In the latter case, the average squared error is shown to generally trend down as system size increases, but this trend is not observed when the TPQ states are formed from poor approximations to Haar-random states. This aligns with expectations from Ref.~\cite{Sugiura_2013}. For low $N$, the error from utilizing a single TPQ state may become large.  Fig. \ref{fig:errorfig}b demonstrates that averaging over the results from multiple TPQ states can reduce such error. In this example, the observable of interest is the thermal energy as a function of inverse temperature $\beta$, and the simulated system is a 6-spin Heisenberg model. 

\bibliography{references}

\section*{Acknowledgments} This research was supported by the Office of Science, Office of Advanced Scientific Computing Research Accelerated Research for Quantum Computing Program of the U.S. Department of Energy under Contract No. DE-AC02-05CH11231.  This research used resources of the National Energy Research Scientific Computing Center (NERSC), a U.S. Department of Energy Office of Science User Facility located at Lawrence Berkeley National Laboratory, operated under Contract No. DE-AC02-05CH11231.

\section*{Author Contributions}
C.P., L.B.O, and W.A.d.J conceived the idea and designed the research.  C.P., D.C., and L.B.O wrote the simulation code and performed simulations.  C.P. and L.B.O wrote the manuscript.  All authors reviewed and edited the manuscript and approved the final manuscript.

\section*{Data Availability Statement}
The datasets used and/or analysed in this work are available from the corresponding author on reasonable request.

\section*{Additional Information}
The authors declare no competing interests.

\end{document}